\documentclass[twocolumn,prb,showpacs]{revtex4}
\usepackage{epsf}

\newcommand{\Av}[1]     {\left\langle #1 \right\rangle}

\newcommand{\eqn}[1]    {(\ref{#1})}
\newcommand{\fig}[1]	{Fig.~\ref{#1}}
\newcommand{\subbox}[1]	{{\mbox{\scriptsize #1}}}

\def\bi         {\begin{itemize}}
\def\ei         {\end{itemize}}
\def\benu	{\begin{enumerate}}
\def\eenu	{\end{enumerate}}
\def\bmat       {\left[ \begin{array}}
\def\emat       {\end{array} \right]}
\def\beq	{\begin{equation}}
\def\eeq	{\end{equation}}
\def\beqn       {\begin{eqnarray*}}
\def\eeqn       {\end{eqnarray*}}
\def\beqa       {\begin{eqnarray}}
\def\eeqa       {\end{eqnarray}}
\def\bquote	{\begin{quote}}
\def\equote	{\end{quote}}
\def\f          {\frac}
\def\bwide	{\begin{widetext}}
\def\ewide	{\end{widetext}}

\def\a          {\alpha}
\def\b          {\beta}

\def\e          {\epsilon}

\def\g          {\gamma}
\def\k          {\kappa}

\def\m          {\mu}

\def\th         {\theta}
\def\ve		{\varepsilon}

\def\x          {\xi}
\def\z		{\zeta}

\def\D          {\Delta}

\def\bk         {{\bf k}}

\def\bq         {{\bf q}}

\def\dag	{\dagger}

\begin{document}
\title{Interplay between parallel and diagonal electronic nematic phases in interacting systems}
\author{Hyeonjin Doh}
\email{hdoh@physics.utoronto.ca}
\affiliation{Department of Physics, University of Toronto, Toronto, 
Ontario M5S 1A7, Canada}
\author{Nir Friedman}
\affiliation{Department of Physics, University of Toronto, Toronto, 
Ontario M5S 1A7, Canada}
\author{Hae-Young Kee}
\email{hykee@physics.utoronto.ca}
\affiliation{Department of Physics, University of Toronto, Toronto, 
Ontario M5S 1A7, Canada}
\date{\today}
\begin{abstract}
An electronic nematic phase can be classified by 
a spontaneously broken discrete rotational symmetry of a host lattice.
In a square lattice, there are two distinct nematic phases.
The parallel nematic phase breaks $x$ and $y$ symmetry,
while the diagonal nematic phase breaks the diagonal $(x+y)$ and
anti-diagonal $(x-y)$ symmetry.
We investigate the interplay between the parallel and diagonal nematic orders 
using mean field theory.
We found that the nematic phases compete with each other,
while they coexist in a finite window of parameter space. 
The quantum critical point between the diagonal nematic and isotropic phases exists,
and its location in a phase diagram depends on the topology of the Fermi surface.
We discuss the implication of our results in the context of neutron scattering and 
Raman spectroscopy measurements on La$_{2-x}$Sr$_x$CuO$_4$.
\end{abstract}

\maketitle
\section{Introduction}
Recently, there has been a great effort to understand intrinsic phases of
a doped Mott insulator in the context of high temperature superconductors.
It has been proposed that quantum fluctuations of a Mott insulator introduced
by hole doping lead to intermediate forms of matter, dubbed as electronic
smectic and nematic phases.\cite{Kivelson98nature,Kivelson03rmp}
In analogy to classical liquid crystals, the smectic phase breaks translational symmetry 
along one direction, while the nematic phase breaks rotational symmetry.

The evidence of such inhomogeneous and/or anisotropic liquid phases has been 
found in strongly correlated electron systems.
\cite{Tranquada95nature,Mori98nature, Lilly99prl, Du99ssc, Cooper02prb, Ando02prl}
In particular,  the clear evidence of a nematic liquid phase has been reported 
in two-dimensional electron gases in magnetic fields in ultra-clean samples.\cite{Lilly99prl}
The observed strong anisotropy of longitudinal resistivity has been explained 
by the onset of a nematic phase at low temperatures.
A recent theoretical study of the nematic phase using a quadrupolar interaction,
$F_2$, has offered non-Fermi liquid behavior in the nematic phase as well as near
the quantum critical point.\cite{Oganesyan01prb}
This is originated from large fluctuations of the overdamped collective modes within the RPA theory.
A non-perturbative approach using higher dimensional bosonization reproduced the quantum critical behavior 
with the dynamical exponent of $z=3$, and verified the non-Fermi liquid behavior in the nematic phase.
\cite{Lawler05cond_mat}

However, it was shown that the nature of phase transition 
of the model with the quadrupolar interaction is quite different when we take into 
account an underlying square lattice.\cite{KeeHY03prb,Khavkine04prb}
On a lattice, a nematic phase can be achieved via a spontaneously broken
point-group symmetry due to interactions between electrons. For example, 
it can break $x$ and $y$ symmetry of a square lattice.
An essential consequence of  nematic order is a deformation of a Fermi surface.
It was shown that the transition from isotropic liquid to the nematic phase
which breaks $x$ and $y$ symmetry of the square lattice
is strongly first order at low temperatures.
The nematic order parameter jumps at the transition to avoid the van Hove singularity,
thus suppressing the Lifshitz transition.
The transition takes place at arbitrarily small attractive quadrupolar interaction 
at the van Hove band filling.
The transition changes to a continuous one at a finite temperature, but is not affected by 
either the next neighbor hopping, $t'$, nor small dispersion in the third direction.

In this paper, we study two distinct nematic phases in the square lattice,
and investigate the interplay between them.
The parallel nematic phase previously studied\cite{Khavkine04prb}
breaks a symmetry between $x$ and $y$, 
while the diagonal nematic phase breaks a symmetry between two diagonal, $(x+y)$
and $(x-y)$ directions.
The order parameter associated with the parallel $(\D)$ and diagonal $(\D')$
nematic phases are defined as follows:
\beqa
\nonumber
\D &=& F_2 \sum_{\bk} (\cos k_x - \cos k_y)\Av{c_\bk^\dag c_\bk^{}},\\
\D' &=& F_2' \sum_{\bk} (2\sin k_x \sin k_y) \Av{c_\bk^\dag c_\bk^{}}.
\label{eq:Nematic_Order_Definitions}
\eeqa
It was shown that $\D'$ is always 0 for a given quadrupolar interaction $F_2$,
which implies that a preferred direction for electron momenta has been selected to
be parallel to the crystal axes.
Here we study the interplay between parallel and diagonal nematic phases using 
a phenomenological model with two different strengths of interactions, $F_2$ and $F_2'$
for the parallel and diagonal nematic orders, respectively.
We find that the transition to the diagonal nematic ordered state
from isotropic liquid phase occurs above a critical value of interaction $F_2'$, 
and it is second order as a function of chemical potential.
The competition between the diagonal and parallel nematic phases leads to 
suppression of the strengths of both phases,
while they coexist in a finite window of chemical potential.

The paper is organized as follows.
We describe the effective model Hamiltonian for the nematic order 
in section \ref{sec:theory}.
The mean field analysis at zero temperature is given 
in section \ref{sec:Results}. 
The effect of $t'$ is also presented. 
We discuss the implication of our results and compare with neutron scattering
 and Raman spectroscopy measurements on La$_{2-x}$Sr$_x$CuO$_4$
in section \ref{sec:Discussion}.
We provide the summary of our findings and future works in the last section.

\section{Hamiltonian for Nematic orders}
\label{sec:theory}
Within a weak-coupling theory,
the instability toward a Fermi surface deformation,
often referred to as Pomeranchuk instability,\cite{Pomeranchuk58jetp}
has been discussed in a Fermi liquid, $t-J$ model, Hubbard model, 
and the extended Hubbard model.\cite{Nilsson05prb,Yamase00jpn,Halboth00prl,Hankevych02prb,Neumayr03prb}
It was shown that a strongly nematic phase (quasi-1D) is stable in a strong coupling limit of 
the two dimensional Emery model.\cite{Kivelson04prb}
A phenomenological model with quadrupolar density interaction in the continuum case
was introduced in Ref.\onlinecite{Oganesyan01prb},
and it was extended to the square lattice in Ref.\onlinecite{KeeHY03prb,Yamase05prb}.

The quadrupolar density interaction 
involves two distinct nematic phases in the square lattice.
For the parallel nematic phase,
the Fermi surface expands along $k_x(k_y)$-axis and shrinks along the
$k_y(k_x)$-axis.
On the other hand,
for the diagonal nematic phase, the Fermi surface expands along $(k_x+k_y)$
and shrinks along the $(k_x-k_y)$ directions (or vice versa).
While the mean field study for the case of $F_2(\bq) = F_2^\prime (\bq)$ showed no
preference of diagonal order,\cite{KeeHY03prb,Khavkine04prb}
the various experiments in cuprates indicates possibility of both parallel
and diagonal fluctuating stripes.\cite{Wakimoto00prb,Fujita02prb,Tassini05prl}
Here we consider the following model Hamiltonian which offers us to study the interplay between the parallel and diagonal nematic phases.
\beqa
\nonumber
\label{eqnhamiltonian}
H &=& \sum_{\bk\sigma} \ve_\bk c_{\bk\sigma}^\dag c_{\bk\sigma}^{} \\
\nonumber
&& - \sum_{\bk\bk'\bq\sigma \sigma^\prime} \left[
F_2(\bq) \z_1(\bk)\z_1(\bk')
+F_2'(\bq) \z_2(\bk)\z_2(\bk')\right] \\
&&\times
c_{\bk+\f{\bq}{2}\sigma}^\dag c_{\bk'-\f{\bq}{2}\sigma^\prime}^\dag
c_{\bk'+\f{\bq}{2}\sigma^\prime}^{} c_{\bk-\f{\bq}{2}\sigma}^{},
\eeqa
where
$F_2 (\bq)$ and $F_2'(\bq)$ are given as follows. 
\beq
F_2(\bq) = \f{F_2}{1+\k q^2},~~~~~
F_2'(\bq) = \f{F_2'}{1+\k' q^2}.
\eeq
Here $\epsilon({\bf k})$, $\z_1(\bk)$, and $\z_2(\bk)$ are  given by
\beqa
\label{eqn:dispersion}
\ve(\bk) &=& -t (\cos k_x + \cos k_y) - 2 t' \cos k_x \cos k_y - \m \\
\z_1(\bk) &=& \cos k_x - \cos k_y \\
\z_2(\bk) &=& 2 \sin k_x \sin k_y. 
\eeqa
The mean field Hamiltonian for the uniform nematic orders is written as
\beq
H_\subbox{mean} =
\sum_\bk \tilde{\ve}_\bk c_\bk^\dag c_\bk^{}
+ \f{|\D|^2}{2F_2} + \f{|\D'|^2}{2F_2'},
\eeq
where
\beq
\tilde{\ve}_\bk = \ve_\bk - \D \z_1(\bk) - \D' \z_2(\bk).
\label{eqn:mean_dispersion}
\eeq
Here $\D$ and $\D'$ measure the strength of the broken $x$ vs. $y$ and $(x+y)$
vs. $(x-y)$ symmetries, respectively.
\beqa
\D &=& F_2 \sum_\bk \left(\cos k_x - \cos k_y\right) \th(-\tilde{\ve})\nonumber\\
\D' &=& F_2' \sum_\bk \left( 2\sin k_x \sin k_y \right) \th(-\tilde{\ve})
\label{eqn:selfconsist}
\eeqa
We compute the free energy using mean field theory, and discuss the phase
transition in the following section.
\section{Phase transition of Nematic orders}
\label{sec:Results}

\subsection{Free energy and Nematic order parameters}
\label{sec:free_energy_n_order}

The mean field free energy at zero temperature is given by 
\beq
F_0(\mu,\D,\D') = \sum_\bk \tilde{\ve}\th(-\tilde{\ve}) 
+ \f{\D^2}{2F_2} + \f{{\D'}^2}{2F_2'}
\label{eqn:FreeEnergy}
\eeq
where $\th(\e)$ is step function. 
Using adaptive 2-dimensional integration,\cite{Henk01prb} we obtain the 
free energy in terms of chemical potential and the nematic orders, $\D$ and $\D'$
Here, we ignored the next-nearest hopping $t'$, 
but we will consider it later in sec.~\ref{sec:nnhopping}.

\begin{figure}[htb]
\epsfxsize=8cm
\epsffile{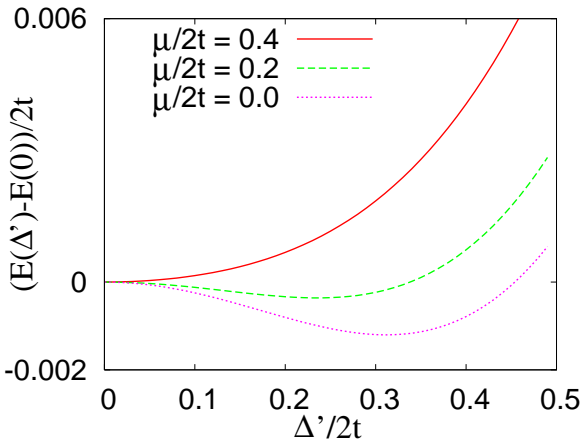}
\caption{
Free energy as a function of  $\D'$ for various values of chemical potential
and $F_2' N_0 = 0.2077$.
The value of free energy has been rescaled to $0$ at $\D'=0$.
\label{fig:SndOrder}
}
\end{figure}

\begin{figure}[htb]
\epsfxsize=8cm
\epsffile{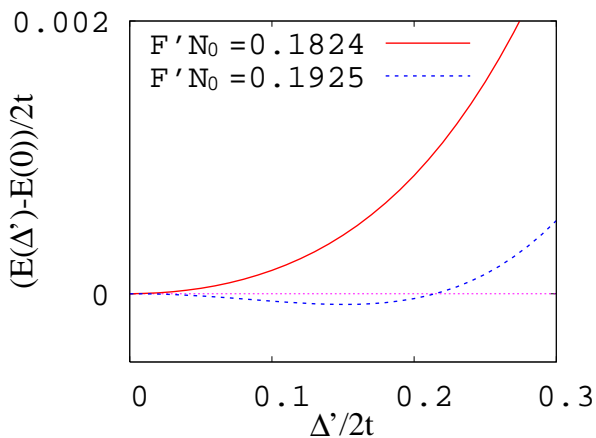}
\caption{
Free energy as a function of  $\D'$ for several values of the interaction, $F_2'$
and $\m =0$.
The free energy has minimum at a finite $\D'$ only when $F_2'$ is larger than
a critical value.
Ihe value of free energy has been rescaled to $0$ at $\D'=0$.
\label{fig:F2pCritic}
}
\end{figure}

To understand the nature of the transition between the diagonal nematic
and isotropic phases, let us first set $F_2=0$.
\fig{fig:SndOrder} shows the free energy as a function of the
diagonal nematic order, $F(\D')$ for several values of chemical potential.
It is clear that the transition from the diagonal nematic phase to the isotropic phase
is second order;
the diagonal order parameter as a function of chemical potential changes continuously.
We found that $\D'$ has finite value only when
$F_2'$ exceeds some critical value $F_{2c}'$ which depends on
the value of $\D$. For the case of $F_2 =0$, $F_{2c}'N_0$ is $0.1876$ 
from our numerical calculation.
\fig{fig:F2pCritic} shows that there is no minimum in the free 
energy except $\D' = 0$ for $F_2'N_0 < F_{2c}'N_0=0.1876$.

\begin{figure*}[hbt]
\epsfxsize=14cm
\epsffile{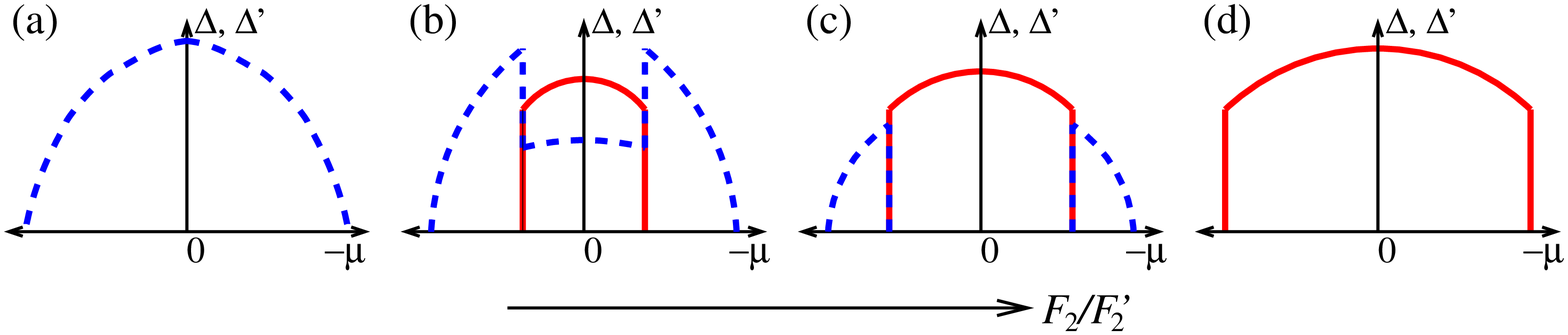}
\caption{
All types of phase diagrams for  $\D$ and $\D'$ as a function
of chemical potential obtained by tuning the ratio between two
interactions $F_2/F_2'$, and $t'=0$.
The solid and dashed lines denote $\D$ and  $\D'$, respectively.
See the main text for the discussion on (a) -(d).
\label{fig:NematicEvolution1}
}
\end{figure*}

We now turn on $F_2$ to understand the interplay between parallel and diagonal nematic orders.
The behaviors of nematic order parameters as one varies $F_2/F_2'$ 
are obtained by solving the self-consistent equation, \eqn{eqn:selfconsist}.
All possible types of phase diagram for $\D$ and $\D'$
are summarized  in \fig{fig:NematicEvolution1}.
As we increase $F_2$, the parallel nematic order suddenly develops near
$\m=0$ where the van Hove singularity exists. This is illustrated in
\fig{fig:NematicEvolution1} (b).
The suppression of diagonal nematic order
due to the development of the parallel nematic order is clearly observed as well.
However, the region of finite $\D'$ does not change as long as $F_2'$ is fixed.
As shown in \fig{fig:NematicEvolution1} (c),
a further increase of $F_2$ leads to a wider region of the parallel
nematic order, which now totally suppresses the diagonal nematic order inside
its territory.
However, it does not take over the whole region of the diagonal
nematic order. A finite $\D'$ region outside the territory of the parallel
nematic order is still found.
A further increase of $F_2$ eventually removes the diagonal nematic phase
in the picture as shown in \fig{fig:NematicEvolution1} (d).

\begin{figure}[htb]
\epsfxsize=7.5cm
\epsffile{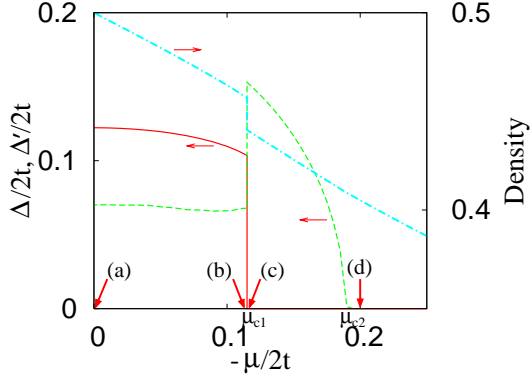}
\caption{
Nematic orders and electron density vs. chemical potential
for $F_2N_0 = 0.1$ and $F_2'N_0 = 0.196$.
The solid and dashed lines denote $\D$ and $\D'$, respectively.
The dot-dashed line denotes the electron density.
The arrows labeled (a) to (d) are the values of chemical potentials used in
\fig{fig:FreeEnergys} and the discussion in the main text.
\label{fig:Nematic1}
}
\end{figure}

\begin{figure}[htbf]
\epsfxsize=8cm
\epsffile{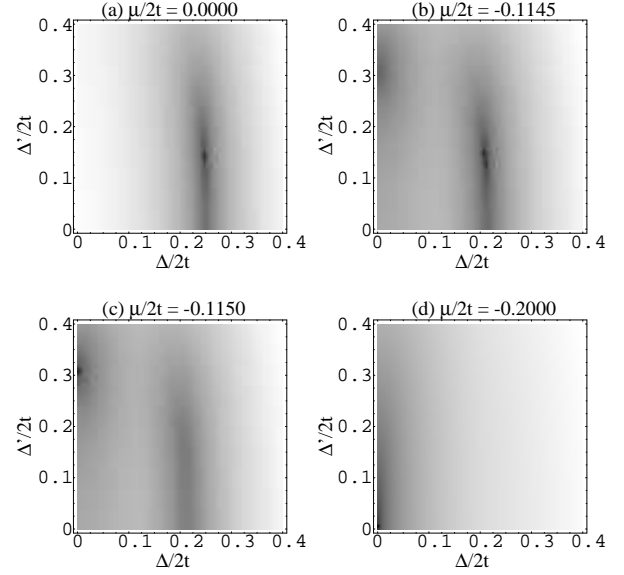}
\caption{
The density plot of free energy as functions of 
$\D$ and $\D'$  for several values of  chemical potentials
indicated by the arrows in \fig{fig:Nematic1}, 
and for given values of interactions, $F_2N_0 = 0.1$ and $F_2'N_0 = 0.196$.
Darker is lower free energy. 
The minimum points for each figure denote the equilibrium values of 
$\D$ and $\D'$. The discussion on (a) - (d) can be found in the main text.
\label{fig:FreeEnergys}
}
\end{figure}

It is worthwhile to emphasize 
the following important features of the nematic orders in the coexistence regime.
The region where the diagonal nematic order is finite is
always wider than that of the parallel nematic order, if it ever exists.
As shown in \fig{fig:Nematic1}, $\D$ has a finite value for 
$ |\m/2t| < 0.115$, and $\D'$ for $ |\m/2t| < 0.19 $.
In other words, the critical chemical potential for $\D'$, $\m_{c2}$ is 
always bigger than that for $\D$, $\m_{c1}$. 
At $\m = \m_{c1}$, $\D$ drops to zero discontinuously  
and the electron density also changes abruptly,
which shows the first order phase transition.
When $\D$ goes to zero at $\m_{c1}$, $\D'$ gets enhanced
from $0.0679$ to $0.1529$ (in unit of $ 2t$). 
We found that the region of $\D$ is also shrunk
due to finite $\D'$, which is further discussed in the following subsection.
As the chemical potential approaches $\m_{c2}$, $\D'$ gradually goes to zero reflecting 
the second order phase transition.
At this transition, the electron density varies continuously and 
shows only the tiny change of its slope.

We show the behavior of free energies around the critical chemical potentials,
to highlight these two distinctive phase transitions at $\m_{c1}$ and $\m_{c2}$ in
\fig{fig:FreeEnergys}.
\fig{fig:FreeEnergys} (a) shows that the minimum of the free energy for $\m/2t=0$ 
occurs at finite $\D$ and $\D'$.
As the chemical potential approaches to the first critical point, $|\m_{c1}/2t|=0.115$ 
another minimum point begins to develop at $\D/2t=0$ and $\D'/2t=0.1529$ as shown
in \fig{fig:FreeEnergys} (b).
At the critical point $\mu_{c1}$, there exist two clear minima as shown in 
\fig{fig:FreeEnergys} (c).
The global minimum of $(\f{\D}{2t},\f{\D'}{2t})$ is changed from
$(0.1035, 0.0679)$ to $(0, 0.1529)$ as $|\m/2t|$ crosses  the critical point, $0.115$.
A further increase of $\m$ finds a unique minimum at  $D=D'=0$ as shown in 
\fig{fig:FreeEnergys} (d).

\subsection{Phenomenological Analysis of two competing orders}

\begin{figure}[htb]
\epsfxsize=9.0cm
\epsffile{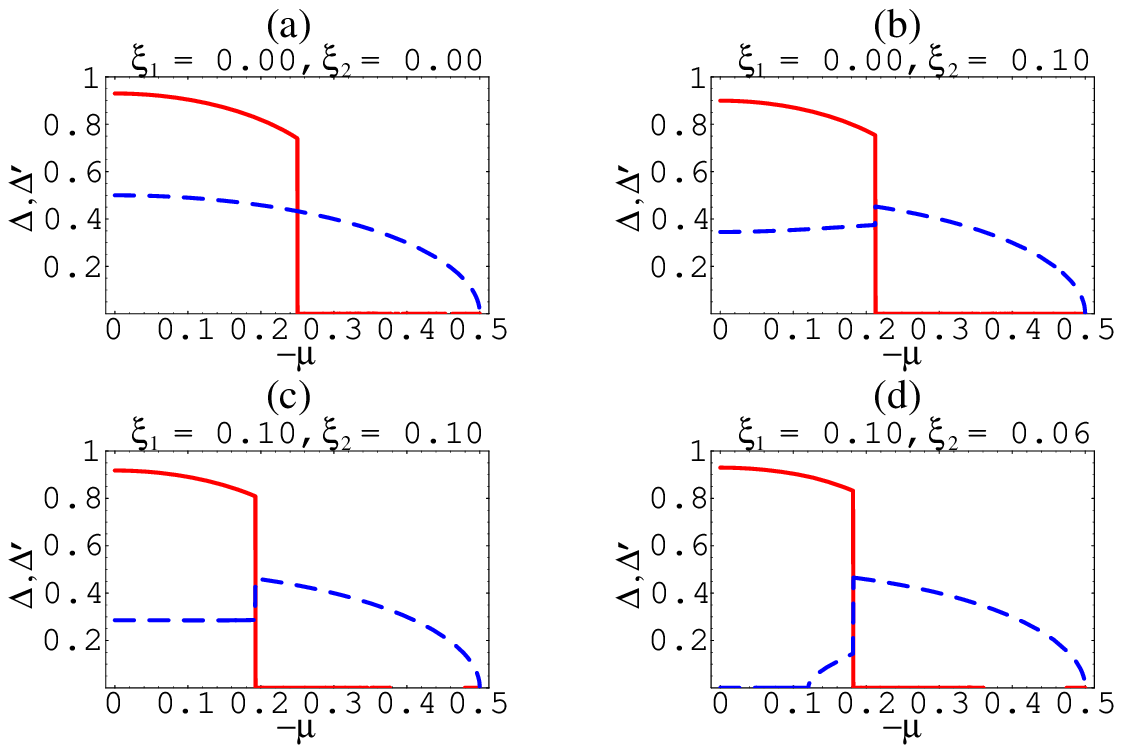}
\caption{
Order parameters determined from the Ginzburg-Landau equation of \eqn{eqn:GL}
for various values of $\x_1$ and $\x_2$.
The other parameters are
$\a = 0.1$,
$\b_0 = 4.0$,
$\g = 1.0$,
$\a_0' = 1.0$,
$\b' = 1.0$,
$\g' = 0.0$,
$\x_3 = 0.0$,
$\m_{c1}=0.25$, and
$\m_{c2}=0.5$.
The solid and dashed lines denote $\D$ and $\D'$, respectively.
\label{fig:GL}
}
\end{figure}

To get an insight on competing two nematic orders,
we analyze the following Ginzburg-Landau (GL) free energy.
\beqa
\nonumber
F_\subbox{GL}(\D,\D')
&=& \f{\a}{2} \D^2 + \f{\b(\m)}{4} \D^4 + \f{\g}{6} \D^6 
\\ &&
 + \f{\a'(\m)}{2} {\D'}^2 + \f{\b'}{4}{\D'}^4 +\f{\g'}{6}{\D'}^6
\label{eqn:GL}
\\ &&
 + \f{\xi_1}{2}{\D}^2{\D'}^2
 + \f{\xi_2}{2}{\D}^4{\D'}^2
 + \f{\xi_3}{2}{\D}^2{\D'}^4.
\nonumber
\eeqa
We expand the free energy in terms of the order parameters up to the 6-th orders, since the parallel
nematic order denoted by $\D$ shows the first order transition.
We set $\g' =\xi_3=0$ for a simplicity,
because different values of $g'$ and $\xi_3$ do not affect our qualitative analysis.
We  introduce positive mutual interaction coefficients
($\x_1 > 0$ and $\x_2 > 0$) between two orders, since the two nematic orders suppress each other.
Here $\a'$($\b$) changes its sign as it crosses the critical chemical potential
$\m_{c2}$($\m_{c1}$), and it is an even function of
the chemical potential due to a particle-hole symmetry.
We consider the following form of $\b(\m)$ and $\a'(\m)$.
\beqa
\b(\m) &=& \b_0 (\m^2 - \m_{c1}^2) - \sqrt{\f{16\a\g}{3}},~~~\mbox{and } \\
\a'(\m) &=& \a_0'(\m^2-\m_{c2}^2).
\eeqa

The order parameters are determined by solving the
following equations, and shown in \fig{fig:GL} for several values of  $\x_1$ and $\x_2$. 
\beqa
\!\!0&\!\!=\!\!&
\left[\a\!\! + \x_1 {\D'}^2\!\!  + \left(\b(\m)+2\x_2{\D'}^2\right)\D^2 
\!\!+ \g \D^4\right]\!\!\D \\
\!\!0&\!\!=\!\!&
\left[\a'(\m)+\x_1 \D^2\!\! +\x_2 \D^4\!\! + \b'{\D'}^2\right]\D' 
\eeqa
%
%
%
The amplitude of $\D$ is not much affected due to the mutual interaction term.
However, the critical chemical potential, $\m_{c1}$ is shifted in 
such a way that the region of the parallel nematic phase becomes narrower.
The modified critical chemical potential, $\tilde{\m}_{c1}$ is given 
by the following equation.
\beq
\tilde{\m}_{c1} =
\sqrt{\m_{c1}-\sqrt{\f{16\a\g}{3\b_0^2}}\left(
\sqrt{1+\f{\x_1{\D'}^2}{\a}}-1\right)-\f{2\x_2{\D'}^2}{\b_0}}.
\eeq
On the other hand, the magnitude of
$\D'$ is  suppressed due to finite $\D$, and it is modified as 
follows.
\beq
\D' = \sqrt{\f{\a_0'(\m_{c2}^2-\m_{}^2)-\x_1\D^2-\x_2\D^4}{\b'}}
\eeq
The critical chemical potential, $\m_{c2}$ is not affected
as long as $\m_{c2} > \tilde{\m}_{c1}$.
Our GL analysis captures the key features of our results presented
in sec.~\ref{sec:free_energy_n_order}.
They suppress each other in qualitatively different ways.
The parallel nematic order suppresses the amplitude of the diagonal nematic order,
while the window of the diagonal nematic order is not affected when they coexist.
On the other hand, the diagonal nematic order shrinks the parallel
nematic order region, but hardly suppresses the amplitude of $\D$.

\subsection{Fermi surface and density of states}
\label{sec:FS_n_DOS}
\begin{figure}[hbt]
\epsfxsize=7.5cm
\epsffile{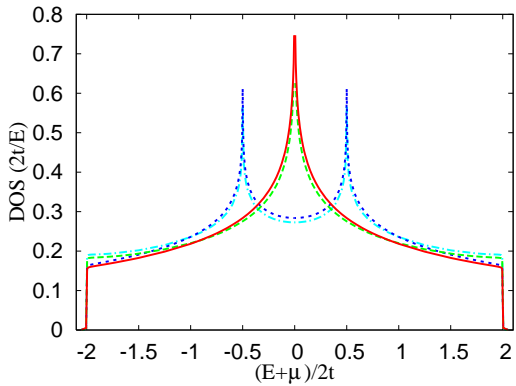}
\caption{
Density of states for various values of $\D$ and $\D'$.
The solid line is for $(\D/(2t),\D'/(2t)) =(0,0)$, 
the dashed  for $(0,0.25)$, the dotted  for $ (0.25,0)$,
and the dot-dashed for $(0.25,0.25)$.
A finite $\D$ splits the van Hove singularity at $E=-\mu$  into two peaks,
while $\D'$ shows the minor effect on the density of states.
\label{fig:dostp0}
}
\end{figure}

\begin{figure}[hbt]
\epsfxsize=7.5cm
\epsffile{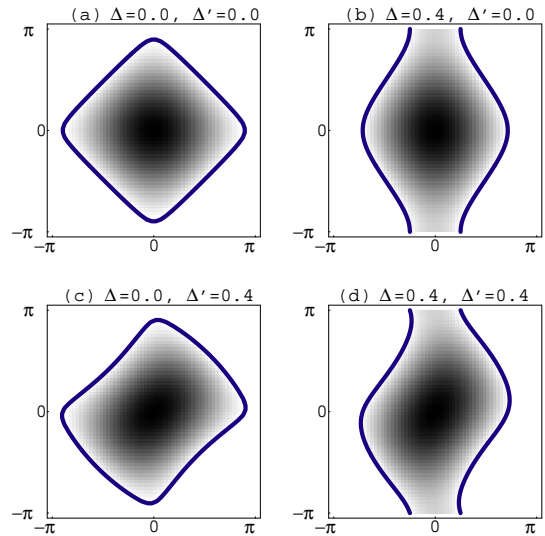}
\caption{
Fermi surface for several values of $\D$ and $\D'$.
The deformation of the original Fermi surface shown in (a)
along the parallel (diagonal) axes of the lattice is due to the parallel (diagonal)
nematic order.  Examples of Fermi surface for the parallel and diagonal nematic order phases
shown in (b) and (c), respectively. 
An example of Fermi surface for finite $\D$ and $\D'$ is shown in (d).
\label{fig:FSshape}
}
\end{figure}

To understand the nature of the phase transition, 
we investigate effects of nematic orders on
the density of states (DOS) and Fermi surface.
The DOS for several values of $\D$ and $\D'$ is shown in in \fig{fig:dostp0}.
Without the nematic orders, DOS has a singularity at $E+\m =0$ 
originated from the van Hove singularity (VHS). 
It was shown that the development of $\D$ (the dotted line)
leads to a dramatic change in DOS. 
The parallel nematic order splits the VHS into two peaks occurring
near the van Hove filling. As a result, the free energy is lowered.\cite{Khavkine04prb}
This feature is inherited from the logarithmic singularity in  the
free energy. (See ref.\onlinecite{Khavkine04prb} for detail.)
On the other hand, $\D'$ is nothing to do with the VHS as shown in \fig{fig:dostp0}.
The peak from VHS is not affected by a development of $\D'$.
The DOS has only minor change due to $\D'$, which is a slight enhancement
near the band edge and suppression near the center of the band.
Therefore, the free energy  develops a minimum continuously from $\D'=0$ to a finite $\D'$,
as one changes $\m$, thus the transition between isotropic and
the diagonal nematic phases is second order.

One of important consequences of the nematic order is
a deformation of the Fermi surface.
The deformation of Fermi surfaces for various values of $\D$ and $\D'$
are shown in \fig{fig:FSshape}.
\fig{fig:FSshape} (a) shows the undeformed Fermi surface to make comparison with (b)-(d).
A finite parallel nematic order, $\D$ squeezes the Fermi surface along a
parallel axis of the lattice as shown in \fig{fig:FSshape} (b). 
For example, it shrinks the Fermi surface in $k_x$-direction
and expands it in $k_y$-direction, or reverse way.
On the other hand, the diagonal nematic order $\D'$ deforms the Fermi surface along
a diagonal axis of the lattice as shown in \fig{fig:FSshape} (c).
For example, it shrinks the Fermi surface in $(k_x-k_y)$-direction
and extend in $(k_x+k_y)$-direction, and vise versa.
The discontinuous change of $\D$ related to the van Hove singularity
leads to a dramatic change in the shape of Fermi surface as shown in 
\fig{fig:FSshape} (b).
On the other hand, the deformation along the diagonal direction develops
continuously, as one changes $\m$.
It is  important to note that
$D'$ does not affect four points on the Fermi surface, $(\pm k_{Fx},0)$ and $(0,\pm k_{Fy})$,
which eventually lead to the van Hove singularity at the van Hove filling and
the formation of the parallel nematic phase inside the diagonal nematic phase.


\subsection{Effects of next-nearest hopping}
\label{sec:nnhopping}

\begin{figure}[htb]
\epsfxsize=8.0cm
\epsffile{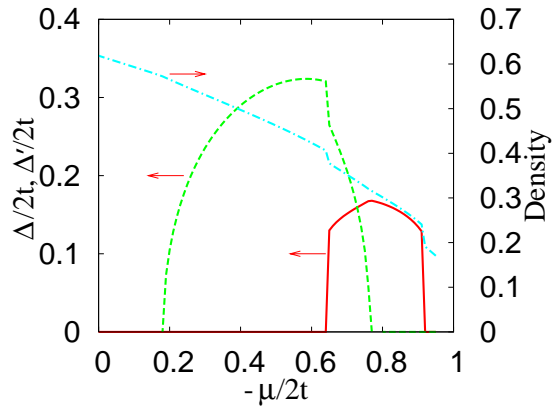}
\caption{
Nematic orders and electron density vs. chemical potential 
for $t'=-0.4t$, $F_2N_0 = 0.1$ and $F_2'N_0=0.196$.
The solid and dashed lines denote $\D$ and $\D'$, respectively.
The dot-dashed line indicates the electron density.
\label{fig:Nematictp1}
}
\end{figure}

\begin{figure*}[hbt]
\epsfxsize=14cm
\epsffile{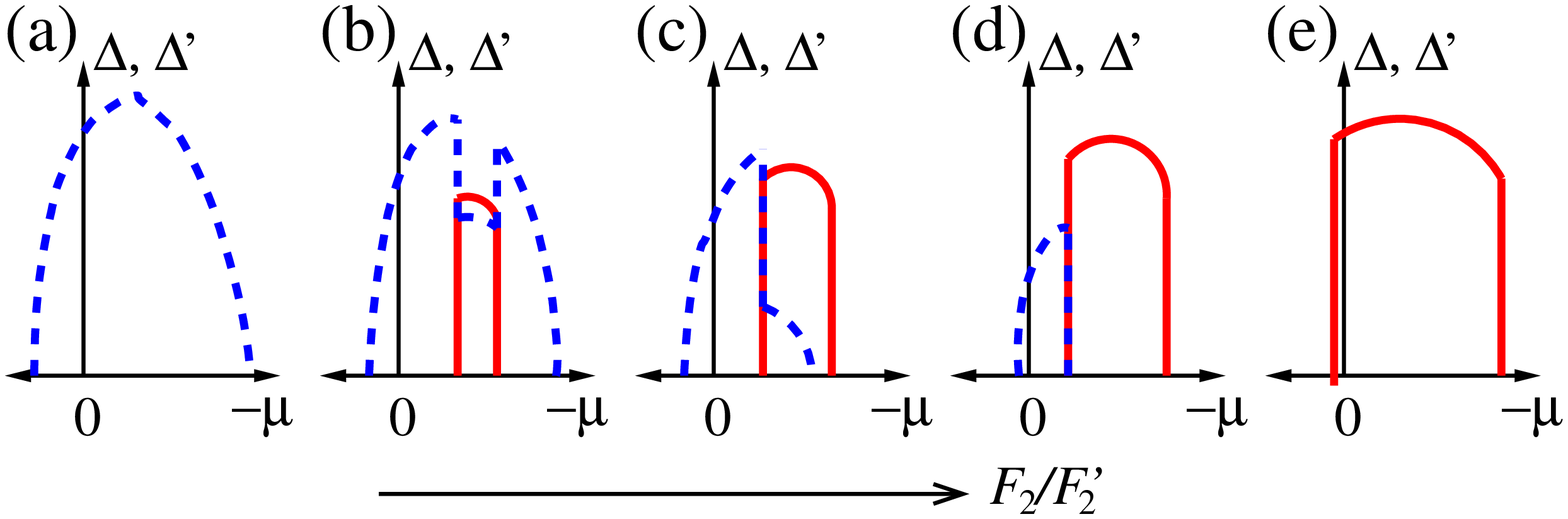}
\caption{
Various types of phase diagrams for $\D$ and $\D'$
with the next-nearest-hopping $t'$. The solid line
is designated for $\D$ and the dashed line for $\D'$.
A negative $t'$ shifts both $\D$ and $\D'$ toward hole-doped region.
(a) $F_2 = 0$ and $F_2' > 0$. 
(b) A small increase of $F_2$ leads to a development of 
the parallel nematic phase inside the diagonal nematic phase region.
(c) A further increase of $F_2$ increases
the parallel nematic phase region, thus suppresses the region of diagonal nematic phase.
(d)  the diagonal nematic order is totally suppressed within the region of parallel nematic phase,
while it still survives outside the region.
(e) The parallel nematic order eventually removes the diagonal nematic order by
expanding its region.
\label{fig:NematicEvolution2}
}
\end{figure*}

In this section, we introduce the next-nearest hopping integral, $t'$ in
\eqn{eqn:dispersion}, which breaks a particle-hole  symmetry.
We found that a negative sign of $t'$ shifts the region of nematic phases toward 
hole-doped region.
However, since the parallel nematic order always appears near a VHS,
while the diagonal nematic order is not directly affected by a VHS,
we expect that the region of parallel nematic and diagonal
nematic phases will be shifted in a slightly different way, as we increase $t'$.
On the other hand, qualitative features such as nature of phase transition
are not affected by a finite $t'$.


\fig{fig:Nematictp1} shows a phase diagram for $t' =-0.4 t$.
The parallel nematic order is shifted more to the hole doped region
than the diagonal nematic order.
At the half-filling, only the diagonal nematic order is finite.
As we increase $|\m|$ (or hole concentration), the coexistence of two nematic phases appears.
A further increase of $|\m|$ leads to a suppression of the diagonal nematic order,
while the parallel nematic phase gets stronger. Finally only the parallel nematic order remains,
which eventually disappears in the phase diagram, as the hole concentration is increased.
The nematic order rotates from the parallel order to the diagonal order, as we increase
hole doping concentration, and two phases coexist near the boundary.
This feature resembles the rotation of charge fluctuations associated with $B_{2g}$ channel
to $B_{1g}$ channel by increasing doping concentration
which was reported in the Raman spectroscopy measurement 
on La$_{2-x}$Sr$_x$CuO$_4$ (LSCO).\cite{Tassini05prl}
The rotation of spin modulation by changing doping concentration
in LSCO was also found in elastic neutron scattering patterns.\cite{Fujita02prb}
Elastic  neutron scattering studies on LSCO showed one-dimensional spin modulation
along the orthorhomic $b$-axis in lower doping concentrations, and another type
of spin modulation parallel to the tetragonal axes in high doping concentrations.
The coexistence of two types of spin modulations near the boundary was also reported.\cite{Fujita02prb}
While direct comparisons to the neutron scattering patterns and/or Raman spectroscopy data
require further theoretical studies on corresponding susceptibilities in the nematic phases,
we expect that the behavior of Fermi surface deformation within our model offers 
a consistent picture compared with the experimental observations.

\fig{fig:NematicEvolution2} summarizes the typical types of
phase diagrams for two nematic orders with the next-nearest hopping.
For a finite negative $t'$,
both $\D(\m)$ and $\D'(\m)$ are shifted to hole-doped region but in
a slightly different way, as we discussed.
When they coexist, the maxima of  two nematic phases do not coincide as shown
in \fig{fig:NematicEvolution1} due to their different $t'$ dependences.
However, the qualitative behaviors of competition between two nematic orders
are similar to the case without $t'$ presented in the sec. III A.

\section{Discussion and Summary}
\label{sec:Discussion}
A quantum analog of classical liquid crystal in terms of broken symmetry
has been discussed in the context of a doped Mott insulator.\cite{Kivelson03rmp}
Among the electronic liquid crystal phases, nematic phases can be viewed as
fluctuating stripes whose segments are fluctuating in time but
oriented in a particular direction, and hence breaks orientational symmetry.

Recent neutron scattering measurements of detwinned YBa$_2$Cu$_3$O$_{7-\delta}$ have indicated 
a possible existence of two dimensional anisotropic liquid crystalline phase
in high temperature cuprates.\cite{Hinkov04nature,Stock04prb,Kao05prb}
Extensive neutron scattering measurements of 
La$_{2-x}$Sr$_x$CuO$_4$ in a wide range of doping have also revealed 
the doping dependence of the static or quasi-static spin ordering in insulating and superconducting phase.\cite{Wakimoto00prb,Fujita02prb}
An interesting observation is that the orientation of the spin modulation
depends on the doping concentration.
It was found that the spin modulation vector is diagonal to the Cu-O bond
in the insulating spin glass phase, 
while inside the superconducting phase it is parallel to the tetragonal axes.
These two types of spin modulation coexist near the boundary between
the insulating and superconducting phases.
Such a one-dimensional nature of the spin correlations is consistent with 
a stripe-like ordering of the holes in the CuO$_2$ planes.\cite{Tranquada95nature}

Inelastic light-scattering spectra of underdoped La$_{2-x}$Sr$_x$CuO$_4$ single crystals showed
additional Drude-like responses in $B_{2g}$ and $B_{1g}$ channels for 
$x=0.02$ to $x=0.10$, respectively.\cite{Tassini05prl} 
This was interpreted as experimental evidence of fluctuating charge stripes 
whose orientation rotates from diagonal to parallel by varying the doping concentration of Sr 
from $x=0.02$ to $x=0.10$.\cite{Tassini05prl}
More extensive studies for various doping concentrations 
will be required to determine the coexistence of two types of
charge modulations.
While  direct comparisons to these experimental data require theoretical studies on
corresponding susceptibilities, we expect that our phenomenological model provides 
a possible explanation of the rotation of the spin and charge modulations observed in
LSCO.

In summary, we have studied the interplay between parallel and diagonal nematic phases using the phenomenological model Hamiltonian within mean field theory.
We found that the parallel and diagonal nematic phases compete each other --
the parallel nematic order suppresses the amplitude of the diagonal nematic order,
while the diagonal nematic order shrinks the window of the parallel nematic order.
However, they still coexist in a finite window of parameter space.
The transition to the parallel nematic phase is strongly first order, while
the diagonal nematic phases shows the continuous transition to the isotropic phase.
Effects of quantum fluctuation of the diagonal nematic order parameter on various quantities
near a quantum critical point  are the subjects of future studies, 
and the single particle self-energy correction due to the fluctuation of a collective mode
will be  presented elsewhere.\cite{Puetter05}

\begin{acknowledgments}
We thank T. P. Devereaux, A. H.  Castro Neto, Y-J Kao, and E. Fradkin for useful discussions.
This work was supported by NSERC of Canada(HD, NF, HYK), Canada Research Chair,
Canadian Institute for Advancded Research, and Alfred P. Sloan
Research Fellowship(HYK).
\end{acknowledgments}

\bibliographystyle{apsrev}

\begin{thebibliography}{27}
\expandafter\ifx\csname natexlab\endcsname\relax\def\natexlab#1{#1}\fi
\expandafter\ifx\csname bibnamefont\endcsname\relax
  \def\bibnamefont#1{#1}\fi
\expandafter\ifx\csname bibfnamefont\endcsname\relax
  \def\bibfnamefont#1{#1}\fi
\expandafter\ifx\csname citenamefont\endcsname\relax
  \def\citenamefont#1{#1}\fi
\expandafter\ifx\csname url\endcsname\relax
  \def\url#1{\texttt{#1}}\fi
\expandafter\ifx\csname urlprefix\endcsname\relax\def\urlprefix{URL }\fi
\providecommand{\bibinfo}[2]{#2}
\providecommand{\eprint}[2][]{\url{#2}}

\bibitem[{\citenamefont{Kivelson et~al.}(1998)\citenamefont{Kivelson, Fradkin,
  and Emery}}]{Kivelson98nature}
\bibinfo{author}{\bibfnamefont{S.~A.} \bibnamefont{Kivelson}},
  \bibinfo{author}{\bibfnamefont{E.}~\bibnamefont{Fradkin}}, \bibnamefont{and}
  \bibinfo{author}{\bibfnamefont{V.~J.} \bibnamefont{Emery}},
  \bibinfo{journal}{Nature (London)} \textbf{\bibinfo{volume}{393}},
  \bibinfo{pages}{550} (\bibinfo{year}{1998}).

\bibitem[{\citenamefont{Kivelson et~al.}(2003)\citenamefont{Kivelson, Fradkin,
  Oganesyan, Bindloss, Tranquada, Kapitulnik, and Howald}}]{Kivelson03rmp}
\bibinfo{author}{\bibfnamefont{S.~A.} \bibnamefont{Kivelson}},
  \bibinfo{author}{\bibfnamefont{E.}~\bibnamefont{Fradkin}},
  \bibinfo{author}{\bibfnamefont{V.}~\bibnamefont{Oganesyan}},
  \bibinfo{author}{\bibfnamefont{I.~P.} \bibnamefont{Bindloss}},
  \bibinfo{author}{\bibfnamefont{J.~M.} \bibnamefont{Tranquada}},
  \bibinfo{author}{\bibfnamefont{A.}~\bibnamefont{Kapitulnik}},
  \bibnamefont{and} \bibinfo{author}{\bibfnamefont{C.}~\bibnamefont{Howald}},
  \bibinfo{journal}{Rev. Mod. Phys.} \textbf{\bibinfo{volume}{75}},
  \bibinfo{pages}{1201} (\bibinfo{year}{2003}).

\bibitem[{\citenamefont{Tranquada et~al.}(1995)\citenamefont{Tranquada,
  Sternlieb, Axe, Nakamura, and Uchida}}]{Tranquada95nature}
\bibinfo{author}{\bibfnamefont{J.}~\bibnamefont{Tranquada}},
  \bibinfo{author}{\bibfnamefont{B.~J.} \bibnamefont{Sternlieb}},
  \bibinfo{author}{\bibfnamefont{J.~D.} \bibnamefont{Axe}},
  \bibinfo{author}{\bibfnamefont{Y.}~\bibnamefont{Nakamura}}, \bibnamefont{and}
  \bibinfo{author}{\bibfnamefont{S.}~\bibnamefont{Uchida}},
  \bibinfo{journal}{Nature (London)} \textbf{\bibinfo{volume}{375}},
  \bibinfo{pages}{561} (\bibinfo{year}{1995}).

\bibitem[{\citenamefont{Mori et~al.}(1998)\citenamefont{Mori, Chen, and
  Cheong}}]{Mori98nature}
\bibinfo{author}{\bibfnamefont{S.}~\bibnamefont{Mori}},
  \bibinfo{author}{\bibfnamefont{C.~H.} \bibnamefont{Chen}}, \bibnamefont{and}
  \bibinfo{author}{\bibfnamefont{S.~W.} \bibnamefont{Cheong}},
  \bibinfo{journal}{Nature (London)} \textbf{\bibinfo{volume}{392}},
  \bibinfo{pages}{473} (\bibinfo{year}{1998}).

\bibitem[{\citenamefont{Lilly et~al.}(1999)\citenamefont{Lilly, Cooper,
  Eisenstein, Pfeiffer, and West}}]{Lilly99prl}
\bibinfo{author}{\bibfnamefont{M.~P.} \bibnamefont{Lilly}},
  \bibinfo{author}{\bibfnamefont{K.~B.} \bibnamefont{Cooper}},
  \bibinfo{author}{\bibfnamefont{J.~P.} \bibnamefont{Eisenstein}},
  \bibinfo{author}{\bibfnamefont{L.~N.} \bibnamefont{Pfeiffer}},
  \bibnamefont{and} \bibinfo{author}{\bibfnamefont{K.~W.} \bibnamefont{West}},
  \bibinfo{journal}{Phys. Rev. Lett.} \textbf{\bibinfo{volume}{82}},
  \bibinfo{pages}{394} (\bibinfo{year}{1999}).

\bibitem[{\citenamefont{Du et~al.}(1999)\citenamefont{Du, Tsui, Stormer,
  Cooper, Pfeiffer, Baldwin, and West}}]{Du99ssc}
\bibinfo{author}{\bibfnamefont{R.~R.} \bibnamefont{Du}},
  \bibinfo{author}{\bibfnamefont{D.~C.} \bibnamefont{Tsui}},
  \bibinfo{author}{\bibfnamefont{H.~L.} \bibnamefont{Stormer}},
  \bibinfo{author}{\bibfnamefont{K.~B.} \bibnamefont{Cooper}},
  \bibinfo{author}{\bibfnamefont{L.~N.} \bibnamefont{Pfeiffer}},
  \bibinfo{author}{\bibfnamefont{K.~W.} \bibnamefont{Baldwin}},
  \bibnamefont{and} \bibinfo{author}{\bibfnamefont{K.~W.} \bibnamefont{West}},
  \bibinfo{journal}{Solid State Commmunication} \textbf{\bibinfo{volume}{109}},
  \bibinfo{pages}{389} (\bibinfo{year}{1999}).

\bibitem[{\citenamefont{Cooper et~al.}(2002)\citenamefont{Cooper, Lilly,
  Eisenstein, Pfeiffer, and West}}]{Cooper02prb}
\bibinfo{author}{\bibfnamefont{K.~B.} \bibnamefont{Cooper}},
  \bibinfo{author}{\bibfnamefont{M.~P.} \bibnamefont{Lilly}},
  \bibinfo{author}{\bibfnamefont{J.~P.} \bibnamefont{Eisenstein}},
  \bibinfo{author}{\bibfnamefont{L.~N.} \bibnamefont{Pfeiffer}},
  \bibnamefont{and} \bibinfo{author}{\bibfnamefont{K.~W.} \bibnamefont{West}},
  \bibinfo{journal}{Phys. Rev. B} \textbf{\bibinfo{volume}{65}},
  \bibinfo{pages}{241313(R)} (\bibinfo{year}{2002}).

\bibitem[{\citenamefont{Ando et~al.}(2002)\citenamefont{Ando, Segawa, Komiya,
  and Lavrov}}]{Ando02prl}
\bibinfo{author}{\bibfnamefont{Y.}~\bibnamefont{Ando}},
  \bibinfo{author}{\bibfnamefont{K.}~\bibnamefont{Segawa}},
  \bibinfo{author}{\bibfnamefont{S.}~\bibnamefont{Komiya}}, \bibnamefont{and}
  \bibinfo{author}{\bibfnamefont{A.~N.} \bibnamefont{Lavrov}},
  \bibinfo{journal}{Phys. Rev. Lett.} \textbf{\bibinfo{volume}{88}},
  \bibinfo{pages}{137005} (\bibinfo{year}{2002}).

\bibitem[{\citenamefont{Oganesyan et~al.}(2001)\citenamefont{Oganesyan,
  Kivelson, and Fradkin}}]{Oganesyan01prb}
\bibinfo{author}{\bibfnamefont{V.}~\bibnamefont{Oganesyan}},
  \bibinfo{author}{\bibfnamefont{S.~A.} \bibnamefont{Kivelson}},
  \bibnamefont{and} \bibinfo{author}{\bibfnamefont{E.}~\bibnamefont{Fradkin}},
  \bibinfo{journal}{Phys. Rev. B} \textbf{\bibinfo{volume}{64}},
  \bibinfo{pages}{195109} (\bibinfo{year}{2001}).

\bibitem[{\citenamefont{Lawler et~al.}(2005)\citenamefont{Lawler, Fernandez,
  Barci, Fradkin, and Oxman}}]{Lawler05cond_mat}
\bibinfo{author}{\bibfnamefont{M.~J.} \bibnamefont{Lawler}},
  \bibinfo{author}{\bibfnamefont{V.}~\bibnamefont{Fernandez}},
  \bibinfo{author}{\bibfnamefont{G.}~\bibnamefont{Barci}},
  \bibinfo{author}{\bibfnamefont{E.}~\bibnamefont{Fradkin}}, \bibnamefont{and}
  \bibinfo{author}{\bibfnamefont{L.}~\bibnamefont{Oxman}}
  (\bibinfo{year}{2005}), \bibinfo{note}{cond-mat/0508747}.

\bibitem[{\citenamefont{Kee et~al.}(2003)\citenamefont{Kee, Kim, and
  Chung}}]{KeeHY03prb}
\bibinfo{author}{\bibfnamefont{H.-Y.} \bibnamefont{Kee}},
  \bibinfo{author}{\bibfnamefont{E.~H.} \bibnamefont{Kim}}, \bibnamefont{and}
  \bibinfo{author}{\bibfnamefont{C.-H.} \bibnamefont{Chung}},
  \bibinfo{journal}{Phys. Rev. B} \textbf{\bibinfo{volume}{68}},
  \bibinfo{pages}{245109} (\bibinfo{year}{2003}).

\bibitem[{\citenamefont{Khavkine et~al.}(2004)\citenamefont{Khavkine, Chung,
  Oganesyan, and Kee}}]{Khavkine04prb}
\bibinfo{author}{\bibfnamefont{I.}~\bibnamefont{Khavkine}},
  \bibinfo{author}{\bibfnamefont{C.-H.} \bibnamefont{Chung}},
  \bibinfo{author}{\bibfnamefont{V.}~\bibnamefont{Oganesyan}},
  \bibnamefont{and} \bibinfo{author}{\bibfnamefont{H.-Y.} \bibnamefont{Kee}},
  \bibinfo{journal}{Phys. Rev. B} \textbf{\bibinfo{volume}{70}},
  \bibinfo{pages}{155110} (\bibinfo{year}{2004}).

\bibitem[{\citenamefont{Pomeranchuk}(1958)}]{Pomeranchuk58jetp}
\bibinfo{author}{\bibfnamefont{I.~J.} \bibnamefont{Pomeranchuk}},
  \bibinfo{journal}{Sov. Phys. JETP} \textbf{\bibinfo{volume}{8}},
  \bibinfo{pages}{361} (\bibinfo{year}{1958}).

\bibitem[{\citenamefont{Halboth and Metzner}(2000)}]{Halboth00prl}
\bibinfo{author}{\bibfnamefont{C.~J.} \bibnamefont{Halboth}} \bibnamefont{and}
  \bibinfo{author}{\bibfnamefont{W.}~\bibnamefont{Metzner}},
  \bibinfo{journal}{Phys. Rev. Lett.} \textbf{\bibinfo{volume}{85}},
  \bibinfo{pages}{5162} (\bibinfo{year}{2000}).


\bibitem[{\citenamefont{Hankevych et~al.}(2002)\citenamefont{Hankevych, Grote,
  and Wegner}}]{Hankevych02prb}
\bibinfo{author}{\bibfnamefont{V.}~\bibnamefont{Hankevych}},
  \bibinfo{author}{\bibfnamefont{I.}~\bibnamefont{Grote}}, \bibnamefont{and}
  \bibinfo{author}{\bibfnamefont{F.}~\bibnamefont{Wegner}},
  \bibinfo{journal}{Phys. Rev. B} \textbf{\bibinfo{volume}{66}},
  \bibinfo{pages}{094516} (\bibinfo{year}{2002}).


\bibitem[{\citenamefont{Neumayr and Metzner}(2003)}]{Neumayr03prb}
\bibinfo{author}{\bibfnamefont{A.}~\bibnamefont{Neumayr}} \bibnamefont{and}
  \bibinfo{author}{\bibfnamefont{W.}~\bibnamefont{Metzner}},
  \bibinfo{journal}{Phys. Rev. B} \textbf{\bibinfo{volume}{67}},
  \bibinfo{pages}{035112} (\bibinfo{year}{2003}).

\bibitem[{\citenamefont{Nilsson and Neto}(2005)}]{Nilsson05prb}
\bibinfo{author}{\bibfnamefont{J.}~\bibnamefont{Nilsson}} \bibnamefont{and}
  \bibinfo{author}{\bibfnamefont{A.~H.~C.} \bibnamefont{Neto}},
  \bibinfo{journal}{Phys. Rev. B} \textbf{\bibinfo{volume}{72}},
  \bibinfo{pages}{195104} (\bibinfo{year}{2005}).

\bibitem[{\citenamefont{Yamase and Kohno}(2000)}]{Yamase00jpn}
\bibinfo{author}{\bibfnamefont{H.}~\bibnamefont{Yamase}} \bibnamefont{and}
  \bibinfo{author}{\bibfnamefont{H.}~\bibnamefont{Kohno}}, \bibinfo{journal}{J.
  Phys. Soc. Jpn.} \textbf{\bibinfo{volume}{69}}, \bibinfo{pages}{2151}
  (\bibinfo{year}{2000}).

\bibitem[{\citenamefont{Kivelson et~al.}(2004)\citenamefont{Kivelson, Fradkin,
  and Geballe}}]{Kivelson04prb}
\bibinfo{author}{\bibfnamefont{S.~A.} \bibnamefont{Kivelson}},
  \bibinfo{author}{\bibfnamefont{E.}~\bibnamefont{Fradkin}}, \bibnamefont{and}
  \bibinfo{author}{\bibfnamefont{T.} \bibnamefont{Geballe}},
  \bibinfo{journal}{Phys. Rev. B} \textbf{\bibinfo{volume}{69}},
  \bibinfo{pages}{144505} (\bibinfo{year}{2004}).


\bibitem[{\citenamefont{Yamase et~al.}(2005)\citenamefont{Yamase, Oganesyan,
  and Metzner}}]{Yamase05prb}
\bibinfo{author}{\bibfnamefont{H.}~\bibnamefont{Yamase}},
  \bibinfo{author}{\bibfnamefont{V.}~\bibnamefont{Oganesyan}},
  \bibnamefont{and} \bibinfo{author}{\bibfnamefont{W.}~\bibnamefont{Metzner}},
  \bibinfo{journal}{Phys. Rev. B} \textbf{\bibinfo{volume}{72}},
  \bibinfo{pages}{35114} (\bibinfo{year}{2005}).

\bibitem[{\citenamefont{Wakimoto et~al.}(2000)\citenamefont{Wakimoto,
  Birgeneau, Kastner, Lee, Erwin, Gehring, Lee, Fujita, Yamada, Endoh
  et~al.}}]{Wakimoto00prb}
\bibinfo{author}{\bibfnamefont{S.}~\bibnamefont{Wakimoto}},
  \bibinfo{author}{\bibfnamefont{R.~J.} \bibnamefont{Birgeneau}},
  \bibinfo{author}{\bibfnamefont{M.~A.} \bibnamefont{Kastner}},
  \bibinfo{author}{\bibfnamefont{Y.~S.} \bibnamefont{Lee}},
  \bibinfo{author}{\bibfnamefont{R.}~\bibnamefont{Erwin}},
  \bibinfo{author}{\bibfnamefont{P.~M.} \bibnamefont{Gehring}},
  \bibinfo{author}{\bibfnamefont{S.~H.} \bibnamefont{Lee}},
  \bibinfo{author}{\bibfnamefont{M.}~\bibnamefont{Fujita}},
  \bibinfo{author}{\bibfnamefont{K.}~\bibnamefont{Yamada}},
  \bibinfo{author}{\bibfnamefont{Y.}~\bibnamefont{Endoh}},
  \bibnamefont{et~al.}, \bibinfo{journal}{Phys. Rev. B}
  \textbf{\bibinfo{volume}{61}}, \bibinfo{pages}{3699} (\bibinfo{year}{2000}).

\bibitem[{\citenamefont{Fujita et~al.}(2002)\citenamefont{Fujita, Yamada,
  Hiraka, Gehring, Lee, Wakimoto, and Shirane}}]{Fujita02prb}
\bibinfo{author}{\bibfnamefont{M.}~\bibnamefont{Fujita}},
  \bibinfo{author}{\bibfnamefont{K.}~\bibnamefont{Yamada}},
  \bibinfo{author}{\bibfnamefont{H.}~\bibnamefont{Hiraka}},
  \bibinfo{author}{\bibfnamefont{P.~M.} \bibnamefont{Gehring}},
  \bibinfo{author}{\bibfnamefont{S.~H.} \bibnamefont{Lee}},
  \bibinfo{author}{\bibfnamefont{S.}~\bibnamefont{Wakimoto}}, \bibnamefont{and}
  \bibinfo{author}{\bibfnamefont{G.}~\bibnamefont{Shirane}},
  \bibinfo{journal}{Phys. Rev. B} \textbf{\bibinfo{volume}{65}},
  \bibinfo{pages}{064505} (\bibinfo{year}{2002}).

\bibitem[{\citenamefont{Tassini et~al.}(2005)\citenamefont{Tassini, Venturini,
  Zhang, Hackl, Kikugawa, and Fujita}}]{Tassini05prl}
\bibinfo{author}{\bibfnamefont{L.}~\bibnamefont{Tassini}},
  \bibinfo{author}{\bibfnamefont{F.}~\bibnamefont{Venturini}},
  \bibinfo{author}{\bibfnamefont{Q.-M.} \bibnamefont{Zhang}},
  \bibinfo{author}{\bibfnamefont{R.}~\bibnamefont{Hackl}},
  \bibinfo{author}{\bibfnamefont{N.}~\bibnamefont{Kikugawa}}, \bibnamefont{and}
  \bibinfo{author}{\bibfnamefont{T.}~\bibnamefont{Fujita}},
  \bibinfo{journal}{Phys. Rev. Lett.} \textbf{\bibinfo{volume}{95}},
  \bibinfo{pages}{117002} (\bibinfo{year}{2005}).

\bibitem[{\citenamefont{Henk}(2001)}]{Henk01prb}
\bibinfo{author}{\bibfnamefont{J.}~\bibnamefont{Henk}}, \bibinfo{journal}{Phys.
  Rev. B} \textbf{\bibinfo{volume}{64}}, \bibinfo{pages}{035412}
  (\bibinfo{year}{2001}).

\bibitem[{\citenamefont{Hinkov et~al.}(2004)\citenamefont{Hinkov, Pailhes,
  bourges, Sidis, Ivanov, Kulakov, Lin, Chen, Bernhard, and
  Keimer}}]{Hinkov04nature}
\bibinfo{author}{\bibfnamefont{V.}~\bibnamefont{Hinkov}},
  \bibinfo{author}{\bibfnamefont{S.}~\bibnamefont{Pailhes}},
  \bibinfo{author}{\bibfnamefont{P.}~\bibnamefont{bourges}},
  \bibinfo{author}{\bibfnamefont{Y.}~\bibnamefont{Sidis}},
  \bibinfo{author}{\bibfnamefont{A.}~\bibnamefont{Ivanov}},
  \bibinfo{author}{\bibfnamefont{A.}~\bibnamefont{Kulakov}},
  \bibinfo{author}{\bibfnamefont{C.~T.} \bibnamefont{Lin}},
  \bibinfo{author}{\bibfnamefont{D.~P.} \bibnamefont{Chen}},
  \bibinfo{author}{\bibfnamefont{C.}~\bibnamefont{Bernhard}}, \bibnamefont{and}
  \bibinfo{author}{\bibfnamefont{B.}~\bibnamefont{Keimer}},
  \bibinfo{journal}{Nature (London)} \textbf{\bibinfo{volume}{430}},
  \bibinfo{pages}{650} (\bibinfo{year}{2004}).

\bibitem[{\citenamefont{Stock et~al.}(2005)\citenamefont{Stock, Buyers, Liang,
  Peets, Tun, Bonn, Hardy, and Birgeneau}}]{Stock04prb}
\bibinfo{author}{\bibfnamefont{C.}~\bibnamefont{Stock}},
  \bibinfo{author}{\bibfnamefont{W.}~\bibnamefont{Buyers}},
  \bibinfo{author}{\bibfnamefont{R.}~\bibnamefont{Liang}},
  \bibinfo{author}{\bibfnamefont{D.}~\bibnamefont{Peets}},
  \bibinfo{author}{\bibfnamefont{Z.}~\bibnamefont{Tun}},
  \bibinfo{author}{\bibfnamefont{D.}~\bibnamefont{Bonn}},
  \bibinfo{author}{\bibfnamefont{W.~N.} \bibnamefont{Hardy}}, \bibnamefont{and}
  \bibinfo{author}{\bibfnamefont{R.~J.} \bibnamefont{Birgeneau}},
  \bibinfo{journal}{Phys. Rev. B} \textbf{\bibinfo{volume}{69}},
  \bibinfo{pages}{014502} (\bibinfo{year}{2004}).

\bibitem[{\citenamefont{Kao and Kee}(2005)}]{Kao05prb}
\bibinfo{author}{\bibfnamefont{Y.-J.} \bibnamefont{Kao}} \bibnamefont{and}
  \bibinfo{author}{\bibfnamefont{H.-Y.} \bibnamefont{Kee}},
  \bibinfo{journal}{Phys. Rev. B} \textbf{\bibinfo{volume}{72}},
  \bibinfo{pages}{024502} (\bibinfo{year}{2005}).

\bibitem[{\citenamefont{Puetter and {\it et al}}(2005)}]{Puetter05}
\bibinfo{author}{\bibfnamefont{C.}~\bibnamefont{Puetter}} \bibnamefont{and}
  \bibinfo{author}{\bibnamefont{{\it et al}}} (\bibinfo{year}{2005}),
  \bibinfo{note}{unpublished}.

\end{thebibliography}

\end{document}